\documentclass[lettersize,journal]{IEEEtran}
\usepackage{amsmath,amsfonts}
\usepackage{algorithmic}
\usepackage{algorithm}
\usepackage[table]{xcolor}
\usepackage{array}
\usepackage[caption=false,font=normalsize,labelfont=sf,textfont=sf]{subfig}
\usepackage{textcomp}
\usepackage{stfloats}
\usepackage{makecell}
\usepackage[hidelinks]{hyperref} 
\usepackage{verbatim}
\usepackage{graphicx}
\usepackage{color}
\usepackage{calc}
\usepackage{makecell}
\usepackage{cite}
\usepackage{enumitem}
\usepackage{multirow} % 用于 \multirow
\usepackage{tabularx}  % 用于 tabularx 和 X 列
\usepackage{multirow}
\usepackage{amssymb}
\usepackage{pifont}
\usepackage{bbding}
\begin{document}

\title{Toward Integrated Air-Ground Computing and Communications: A Synergy of Computing Power Networks and Low-Altitude Economy Network}

\author{Yan Sun, Yinqiu Liu, Shaoyong Guo*, Ruichen Zhang, Jiacheng Wang, Feng Qi, Xuesong Qiu,~\IEEEmembership{Senior Member,~IEEE,} and Dusit Niyato,~\IEEEmembership{Fellow,~IEEE} 

\thanks{This work was supported by the National Natural Science Foundation of China (62322103). (*Corresponding author: Shaoyong Guo)}% <-this % stops a space
\thanks{Yan Sun, Shaoyong Guo, Feng Qi and Xuesong Qiu are with the State Key Laboratory of Networking and Switching Technology, Beijing University of Posts and Telecommunications, Beijing, 100876, China (E-mail: \{sunyan79, syguo, qifeng, xsqiu\}@bupt.edu.cn).}
\thanks{Yinqiu Liu, Ruichen Zhang, Jiacheng Wang and Dusit Niyato are with the College of Computing and Data Science, Nanyang Technological University, 639798, Singapore (E-mail: yinqiu001@e.ntu.edu.sg, ruichen.zhang@ntu.edu.sg, jcwang\_cq@foxmail.com, DNIYATO@ntu.edu.sg).}

}

\maketitle

\begin{abstract}
With the rapid rise of the Low-Altitude Economy (LAE), the demand for intelligent processing and real-time response in services such as aerial traffic, emergency communications, and environmental monitoring continues to grow. Meanwhile, the Computing Power Network (CPN) aims to integrate global computing resources and perform on-demand scheduling to efficiently handle services from diverse sources. However, it is limited by static deployment and limited adaptability. In this paper, we analyze the complementary relationship between LAE and CPN and propose a novel air–ground collaborative intelligent service provision with an agentification paradigm. Through synergy between LAE and CPNs, computing and communication services are jointly scheduled and collaboratively optimized to enhance the execution efficiency of low-altitude services and improve the flexibility of CPNs. It also integrates LAE's strengths in aerial sensing, mobile coverage, and dynamic communication links, forming a cloud–edge–air collaborative framework. Hence, we review the characteristics and limitations of both LAE and CPN and explore how they can cooperate to overcome these limitations. Then we demonstrate the flexibility of the integrated CPN and LAE framework through a case study. Finally, we summarize the key challenges in constructing an integrated air–ground computing and communication system and discuss future research directions toward emerging technologies.
\end{abstract}


\section{Introduction}
With the rapid rise of Low-Altitude Economy (LAE), diverse applications based on Uncrewed Aerial Vehicles (UAVs) have emerged across various fields, including autonomous air transportation, aerial sensing, and emergency communications \cite{ref1}. The prosperity of LAE not only promotes the efficient utilization of airspace resources but also drives an urgent demand for intelligent communication, real-time computing, and dynamic scheduling. However, existing LAE systems still face many challenges \cite{ref2}. On the one hand, the computing and storage capacities of low-altitude nodes are limited, resulting in high response latency and difficulties in supporting complex intelligent services. Additionally, the LAE network environment is highly dynamic, and the heterogeneous requirements of different services in terms of latency, bandwidth, and energy consumption make traditional service planning and processing models insufficiently generalizable, hindering their ability to simultaneously meet multidimensional service demands and thus affecting system stability and service quality.

Meanwhile, the Computing Power Network (CPN), as a new paradigm of network infrastructure, introduces a computing-centric network organization model aimed at aggregating global computing power and enabling unified, on-demand scheduling of resources across cloud, edge, and terminal domains \cite{ref3}. Through key technologies such as computing power abstraction, computation-aware routing, in-network computing, and programmable control, the CPN provides a systematic foundation for realizing Computing-as-a-Service (CaaS), endowing the network with intelligence, flexibility, and evolvability. However, due to the static deployment of CPN nodes, its adaptability to dynamic environments remains limited, and its resource topology lacks elasticity\cite{ref4}.

From the above analysis of the capabilities and limitations of both CPN and LAE, we observe that they possess inherent complementarity in terms of resource attributes and operational mechanisms. Specifically, LAE features wide-area mobility and real-time sensing capabilities, enabling flexible coverage and information acquisition in dynamic environments, while CPN provides powerful global computing scheduling and intelligent optimization capabilities, enabling unified management and intelligent coordination across multi-domain resources\cite{ref1,ref3}. When tightly integrated, CPN can offer LAE cross-space and cross-domain intelligent computing and global optimization support, enhancing the intelligent computing capacity of low-altitude nodes. Additionally, the mobility and aerial distribution of LAE can expand the coverage and adaptability of CPN, equipping it with scalable support for dynamic networks. This computation–communication complementarity and air–ground collaboration provide a new technological pathway for constructing an integrated intelligent air–ground network system.

To demonstrate the aforementioned complementary, in this paper, we propose an integrated air–ground computing and communication framework. 
It establishes unified intelligent service provisions by combining the computation–network collaboration capabilities of CPNs with the aerial sensing and flexible coverage of LAE. 
In detail, the CPN leverages its global computing power scheduling, programmable network management, and in-network computing capabilities to provide LAE with high-performance intelligent training, data processing, task offloading, and joint communication–computation optimization\cite{ref3}. Meanwhile, LAE nodes, through their mobility and aerial communication features, enhance the scalability and dynamic adaptability of the CPN, enabling mobile computing extension, task migration, aerial relaying, and distributed in-network computing\cite{ref2}. Through this bidirectional collaboration mechanism, the air–ground system forms an integrated cloud–edge–air intelligent network infrastructure that supports flexible computing resource allocation, ubiquitous communication connectivity, and sustainable intelligent services. The main contributions of this paper are as follows.
\begin{itemize}
    \item We review the capabilities and limitations of both LAE and CPN and analyze how they can collaborate to overcome each other's constraints.
    \item We explore the mutual assistance mechanisms between CPN and LAE, illustrating how their unique strengths can be leveraged to compensate for each other’s weaknesses. Our review provides actionable design principles that leverage air-ground synergy to achieve two key advantages: 1) enhancing the execution efficiency and intelligence of LAE services by offloading computation to the CPN, and 2) improving the flexibility and service coverage of the CPN by using LAE nodes as mobile relays and computing extensions.
    \item We present an agentification paradigm-based air–ground collaborative service case study, which concretely demonstrates how the collaborative mechanism between LAE and CPN realizes the above two key advantages. On one hand, LAE nodes act as agile aerial extensions, enhancing the responsiveness and adaptability of the CPN. On the other hand, CPN provides stable and powerful computational support, ensuring a high service success rate for aerial tasks.
\end{itemize}

\section{Overview of Low-Altitude Economy and Computing Power Network}
In this section, we first provide an overview of LAE and CPN, including their unique capabilities and challenges. 
Then, we analyze the new capabilities that can emerge from their integration.

\subsection{Low-Altitude Economy}
As the latest convergence of intelligent services and aerial transportation, LAE has recently attracted significant attention. 
It leverages low-altitude airspace (typically below 1000 meters) to develop and deliver a variety of intelligent aerial services. LAE aims to enable safe, efficient, and intelligent low-altitude activities. Typical application scenarios include urban logistics and parcel delivery, low-altitude inspection, and environmental sensing\cite{ref1}. Driven by 5G/6G communications, edge and cloud computing, and artificial intelligence, LAE is evolving from isolated UAV applications to a highly networked and service-oriented ecosystem. In this emerging paradigm, an integration of computing and communication resources enables dynamic task allocation, real-time situational awareness, and collaborative decision-making among distributed aerial and ground entities. Meanwhile, the openness of low-altitude airspace management and the deployment of unified digital platforms have facilitated large-scale commercial operations and enhanced coordination among air traffic control authorities, service providers, and users\cite{ref5}.

At the same time, numerous studies have focused on LAE network management. For example, the authors in \cite{ref11} proposed an UAV-assisted low-altitude edge intelligence network management framework, in which UAVs act as edge computing nodes to jointly optimize task offloading decisions, resource allocation, and UAV trajectories, thereby addressing the trade-off between minimizing task delay and reducing overall network energy consumption. The authors in \cite{ref12} proposed a multi-agent collaborative task allocation optimization framework for LAE networks, which leverages queueing theory and a multi-agent actor–critic architecture to optimize resource efficiency with respect to long-term computing and communication latency and throughput.

\subsection{Computing Power Networks}
The CPN aims to unify the management of global computing resources and perform on-demand scheduling, integrating distributed and centralized computing with intelligent network control into a single service-oriented system. Unlike traditional communication networks that primarily focus on data transmission, CPN is computing-centric, expanding network functionality to include the discovery, scheduling, and orchestration of heterogeneous computing resources across cloud, edge, and end-device domains\cite{ref3}. Through these capabilities, CPN not only improves resource utilization efficiency but also supports latency-sensitive and computation-intensive applications, including autonomous driving, immersive media, and large-scale AI model deployment. By bridging the gap between computing power and network connectivity, CPN ensures that services can access the right computational resources anytime and anywhere\cite{ref6}. However, current CPN infrastructure relies heavily on static, ground-based nodes. This fixed deployment creates a significant gap when supporting emerging applications that are inherently mobile and dynamic, such as autonomous cars and UAV swarms.

Recently, CPNs have been increasingly integrated with LAE networks to leverage the strengths of both while compensating for each other’s limitations. Through efficient collaboration between distributed and centralized computing resources, CPNs possess strong capabilities in data collection and processing, enabling them to leverage ubiquitous computing resources to assist LAE nodes in handling intelligent communication tasks. For example, the authors in \cite{ref9} proposed a joint task offloading and UAV trajectory optimization framework, where edge CPN nodes collect information from both UAVs and the environment to determine optimized strategies. Given the diversity of LAE scenarios and the lack of model generalization, CPNs can utilize high-performance data centers to perform transfer learning and domain knowledge generalization for AI models. For example, the authors in \cite{ref7} proposed a CPN-enabled multi-UAV collaborative knowledge-sharing framework, where individual UAVs collect data from different environments and train local models. A centralized computing power center then aggregates these datasets and performs transfer learning to train general intelligent models.

\begin{figure*}[!t]
\centering
\includegraphics[width=5.5in]{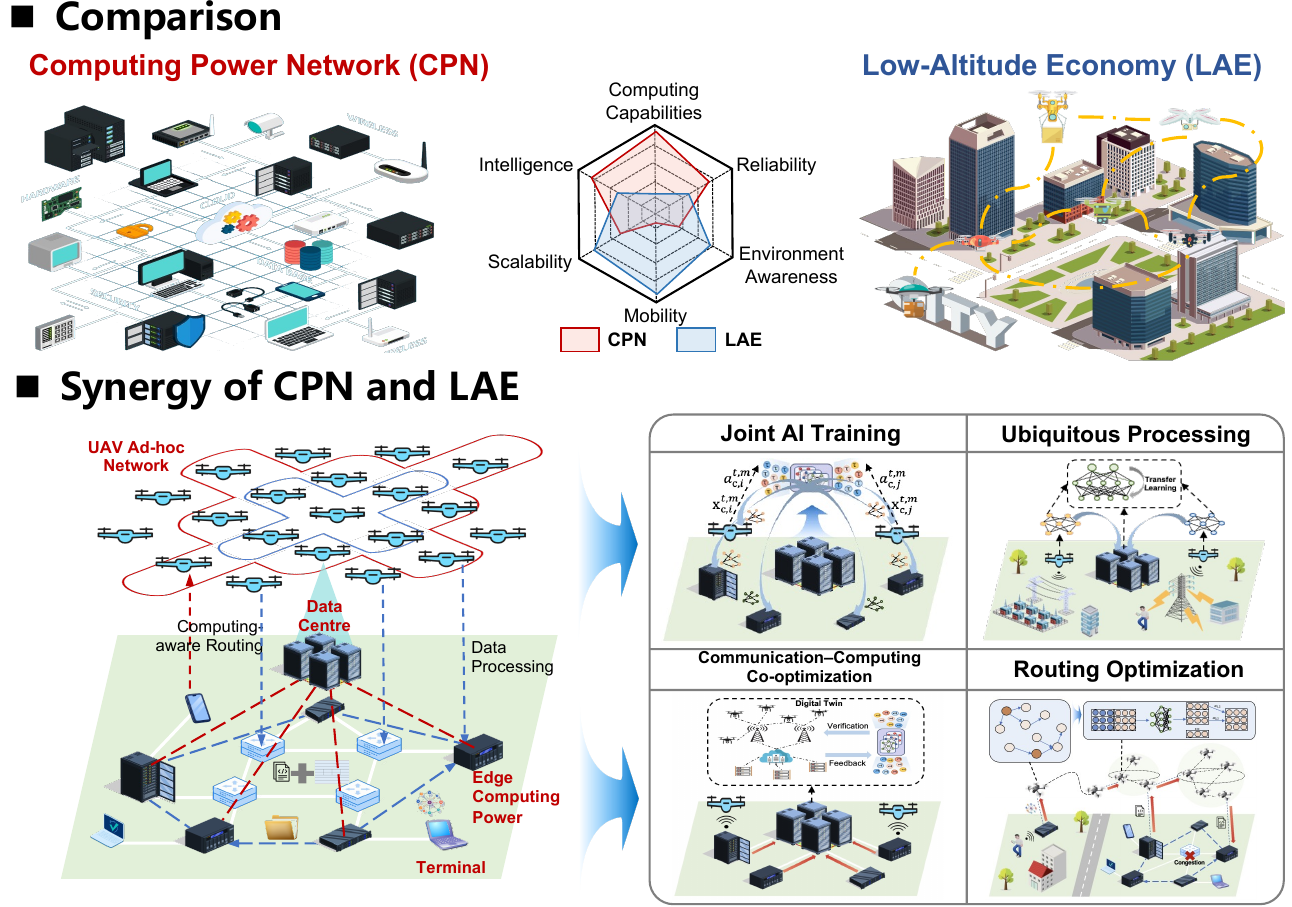}
\caption{LAE features wide-area mobility and real-time sensing capabilities, enabling flexible coverage and information acquisition in dynamic environments, while CPN provides powerful global computing scheduling and intelligent optimization capabilities, enabling unified management and intelligent coordination across multi-domain resources. They can leverage their unique capabilities to assist each other.}
\vspace{-4mm}
\end{figure*}
\subsection{Challenges in CPN and LAE and Their Complementarity}
\subsubsection{Challenges in LAE} The deployment of LAE nodes still faces major challenges arising from limited node capabilities and network diversity\cite{ref2}. Moreover, the high mobility and heterogeneity of LAE networks hinder the generalization of LAE intelligence, as lightweight models cannot adapt well across diverse environments with varying data distributions and service demands\cite{ref7}. The resulting complexity in large-scale, mixed-data transmissions further complicates link optimization and resource allocation, highlighting the need for efficient data classification, intent recognition, and differentiated service provisioning.
\subsubsection{Challenges in CPN}
The deployment of CPN faces significant challenges arising from its infrastructure limitations and the dynamic nature of service demands. First, during periods of high service concurrency, computing nodes may suffer from overload and congestion, leading to increased latency and degraded service continuity\cite{ref4}. The static deployment of ground-based computing resources also limits CPN's ability to dynamically balance workloads across heterogeneous environments. Furthermore, the coverage of CPN remains constrained by its fixed network topology\cite{ref8}. Finally, the limited in-network computing capability of CPN hinders its ability to preprocess and optimize data flows during transmission, resulting in inefficient routing and delayed service responses.

\subsubsection{Complementarity}
Building upon the above analysis of the characteristics and challenges of CPN and LAE, we summarize how the two systems complement each other to achieve better overall performance, as illustrated in Fig. 1. On the CPN-for-LAE side, LAE nodes often lack sufficient computational capabilities, whereas CPN can act as their computing backbone, enabling LAE nodes to offload computation-intensive tasks to ground-based CPN nodes\cite{ref7}. In addition, given the large-scale and heterogeneous data transmission demands in LAE networks, CPN can provide backend computing support for efficient data classification, intent recognition, and differentiated service configuration. On the LAE-for-CPN side, when CPN nodes become overloaded under high service concurrency, LAE nodes can act as mobile or temporary computing relays. Furthermore, LAE nodes can rapidly reach uncovered or hard-to-access areas, extending CPN’s computational services seamlessly into those regions and providing on-demand computing support for remote or mobile users\cite{ref1}.

\section{The Synergy of CPN and LAE}

\begin{table*}[htbp]
\centering
\caption{A Summary of Related Works in the Synergy of Computing Power Network (CPN) and Low-Altitude Economy (LAE) Network}
\includegraphics[width=6in]{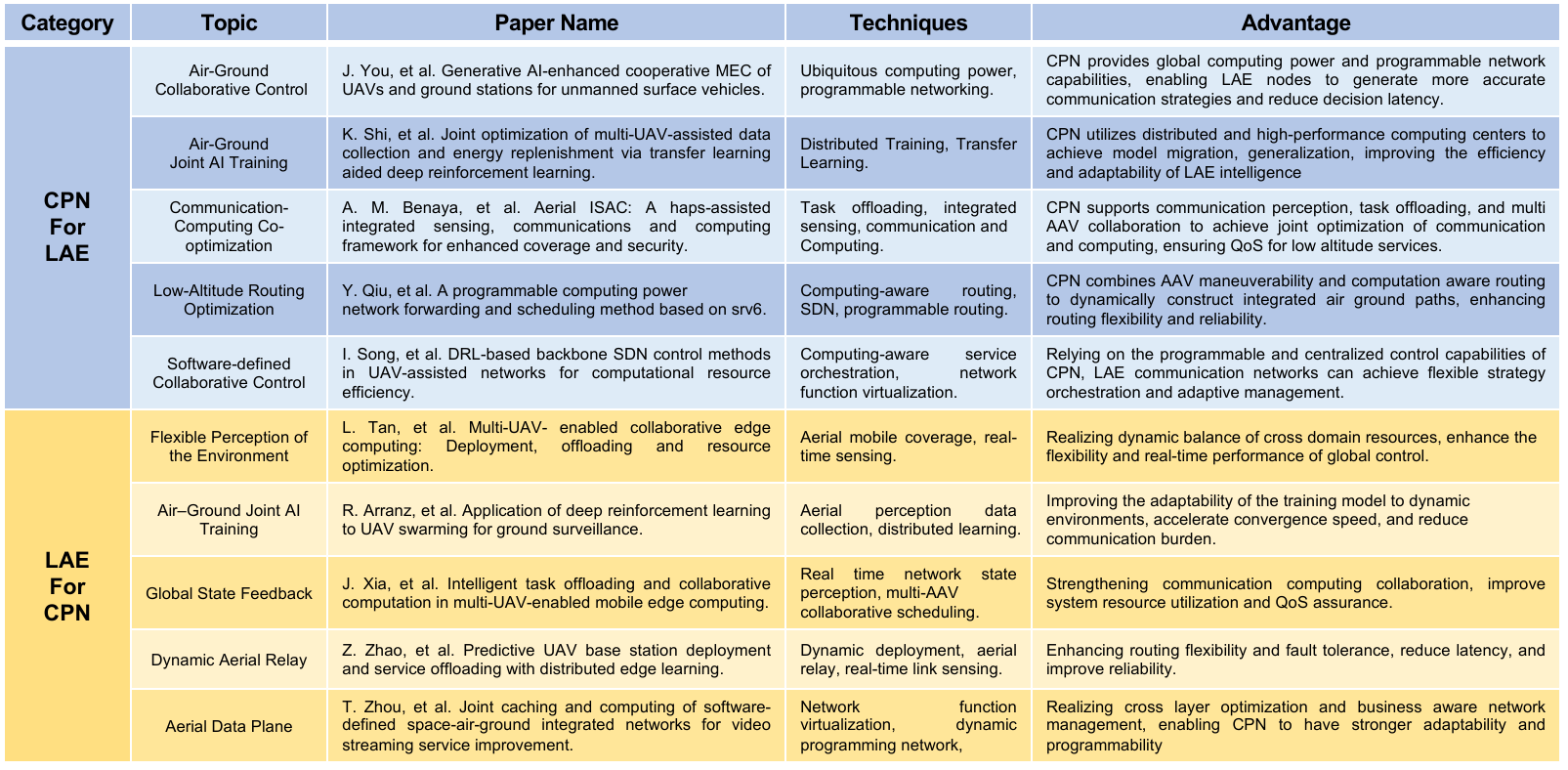}
\vspace{-4mm}
\end{table*}

In this section, we explore the synergy of CPN and LAE in air–ground collaborative control, joint AI training, communication–computing co-optimization, low-altitude routing optimization, and software-defined collaborative control. 
We summarize some of the related works in Table I.

\subsection{CPN for LAE}
\subsubsection{Air–Ground Collaborative Control}
Given the limited computing capability of LAE nodes, CPN can leverage its ubiquitous computing nodes to process multimodal data from the LAE network. Meanwhile, CPN can utilize centralized data centers to integrate these data for executing complex coordination tasks among LAE nodes. This approach alleviates the computational burden of decision-making on resource-constrained LAE nodes by providing them with optimized high-level commands and trajectories, enabling more sophisticated and coordinated autonomous behaviors\cite{ref13}.

\subsubsection{Air–Ground Joint AI Training}
The limited computing and storage capacities of LAE nodes prevent them from training complex AI models locally. CPN can leverage distributed edge computing resources to collect data from LAE nodes and utilize data centers to provide the high-performance computing infrastructure required for centralized model training. By aggregating the large-scale datasets gathered by LAE swarms, CPN enables the training of sophisticated models for tasks such as object detection, semantic segmentation, and navigation\cite{ref7}. Once trained, these models can be optimized, compressed, and deployed back to LAE nodes, following the paradigm of centralized training and distributed inference.

\subsubsection{Communication–Computing Co-optimization}
The LAE nodes lack comprehensive awareness of computing resources, making it difficult to develop scheduling and optimization strategies that simultaneously meet communication and computation demands. This limitation often results in inefficient task offloading and resource allocation. CPN, on the other hand, can leverage its global computing power to collect information about both communication and computation resources, enabling joint optimization through task offloading and multi-node collaborative scheduling. This ensures efficient utilization of resources and guarantees the quality of low-altitude services.

\subsubsection{Low-Altitude Routing Optimization}
In LAE networks, a large volume of heterogeneous service data is transmitted simultaneously, with each service having distinct requirements. However, LAE nodes often struggle to make appropriate routing decisions that account for the characteristics of different data types. CPN can address this challenge by jointly considering real-time communication link quality, computational load, and available resources to make more sophisticated routing decisions. When a service request is initiated, CPN evaluates the specific requirements of the service in terms of bandwidth, latency, and computing capability. It then applies computing-aware routing techniques to determine an optimal service path, which intelligently routes data through specific nodes capable of performing intermediate processing\cite{ref3}.

\subsubsection{Software-defined Collaborative Control}
Acting as a centralized Software-Defined Networking (SDN) controller, the CPN can flexibly manage the air-to-ground and air-to-air communication networks\cite{ref14}. It can dynamically reconfigure the network topology, allocate bandwidth, and manage data flows based on changing mission priorities. For example, it can prioritize a video stream from an LAE that has identified a critical target, or establish a multi-hop relay path through other LAEs to extend communication range. This software-defined approach provides unprecedented flexibility, resilience, and efficiency for the LAE communication network.

\subsection{LAE for CPN}
\subsubsection{Flexible Perception of the Environment}
In CPNs, nodes are typically fixed and lack the flexibility to perceive global environmental dynamics, which can result in decision-making processes that miss critical contextual information. LAE nodes can provide real-time, high-resolution sensing data from advantageous aerial perspectives, offering timely and accurate information for CPN situational awareness and decision-making. They also serve as physical-world executors of high-level CPN commands, thereby closing the perception–action loop\cite{ref1}.

\subsubsection{Air–Ground Joint AI Training}
LAE nodes can operate in diverse and dynamic environments, collecting rich real-world datasets that are essential for training robust and generalizable AI models on the CPN. In addition, LAE nodes can perform initial data filtering, preprocessing, or annotation at the edge, reducing the volume of raw data that needs to be transmitted to the CPN. This conserves bandwidth and enables a more efficient training pipeline. In federated learning scenarios, LAE nodes can even conduct local model training and only transmit model updates, thereby enhancing both data privacy and training efficiency\cite{ref15}.

\subsubsection{Global State Feedback}
To enable the CPN to make effective offloading and resource allocation decisions that satisfy both communication and computing requirements, it requires accurate and real-time feedback from the LAE. LAE nodes provide this by moving through the environment and sensing global information, including resource availability and communication link status. The real-time sensing data collected by LAE nodes is crucial for CPN optimization algorithms, allowing the system to be accurately modeled and enabling more informed and adaptive decision-making.

\subsubsection{Dynamic Aerial Relay}
The CPN may be affected by issues such as network congestion. In such cases, the LAE can serve as a dynamic, on-demand aerial routing relay. LAE nodes can receive data from one computing node and relay it to another, effectively bypassing ground obstacles or congested routes. By forming these aerial data links, clusters of LAE nodes provide resilient extensions to the CPN’s network infrastructure. This ensures robust and continuous connectivity for critical data flows, maintaining the overall stability of the CPN even when ground network performance degrades\cite{ref1}.

\subsubsection{Aerial Data Plane}
Within the software-defined framework, LAE nodes can function as distributed data planes. Critically, they can form dynamic, self-organizing networks that act as aerial relays, extending network coverage to areas without direct CPN-to-ground connectivity. This capability to dynamically establish and reconfigure communication links provides the resilient and scalable service coverage required for complex tasks\cite{ref14}.

\subsection{Lessons Learned}
The CPN provides the distributed computing and control core for the LAE network, offering the large-scale processing power and global optimization capability required to transform LAE from a collection of individual mobile nodes into a coordinated and intelligent aerial swarm\cite{ref6}. Conversely, the LAE network serves as the mobile sensing and execution layer for the CPN, freeing the CPN from static physical infrastructure constraints.
This enhancement transforms the CPN from a fixed, virtualized resource pool into a resilient, adaptive, and physically aware system capable of dynamically extending its services to any environment\cite{ref9}. Ultimately, this synergy lies in fusing the logical and computational intelligence of the CPN with the physical mobility and sensing capabilities of the LAE, thereby creating a network architecture that unifies computational power with dynamic adaptability.

\section{Case Study: Air-Ground Collaborative Intelligent Service Provision with Agentification Paradigm}
\subsection{Background}
Recalling that in the previous section we have clarified the core synergy between LAE and CPN lies in the fusion of CPN’s powerful computational intelligence with LAE’s mobility and sensing capability. In this case study, our aim is to provide a concrete empirical demonstration of that principle. We consider the key challenge of responding to dynamic service hotspots, where static CPN edge nodes are easily overwhelmed by sudden service demands. The core difficulty in addressing this challenge is that traditional rigid CPN control systems cannot perform real-time air–ground coordination to dynamically schedule mobile LAE nodes as aerial relays or computational extensions. 

\begin{figure*}[!t]
\centering
\includegraphics[width=5.5in]{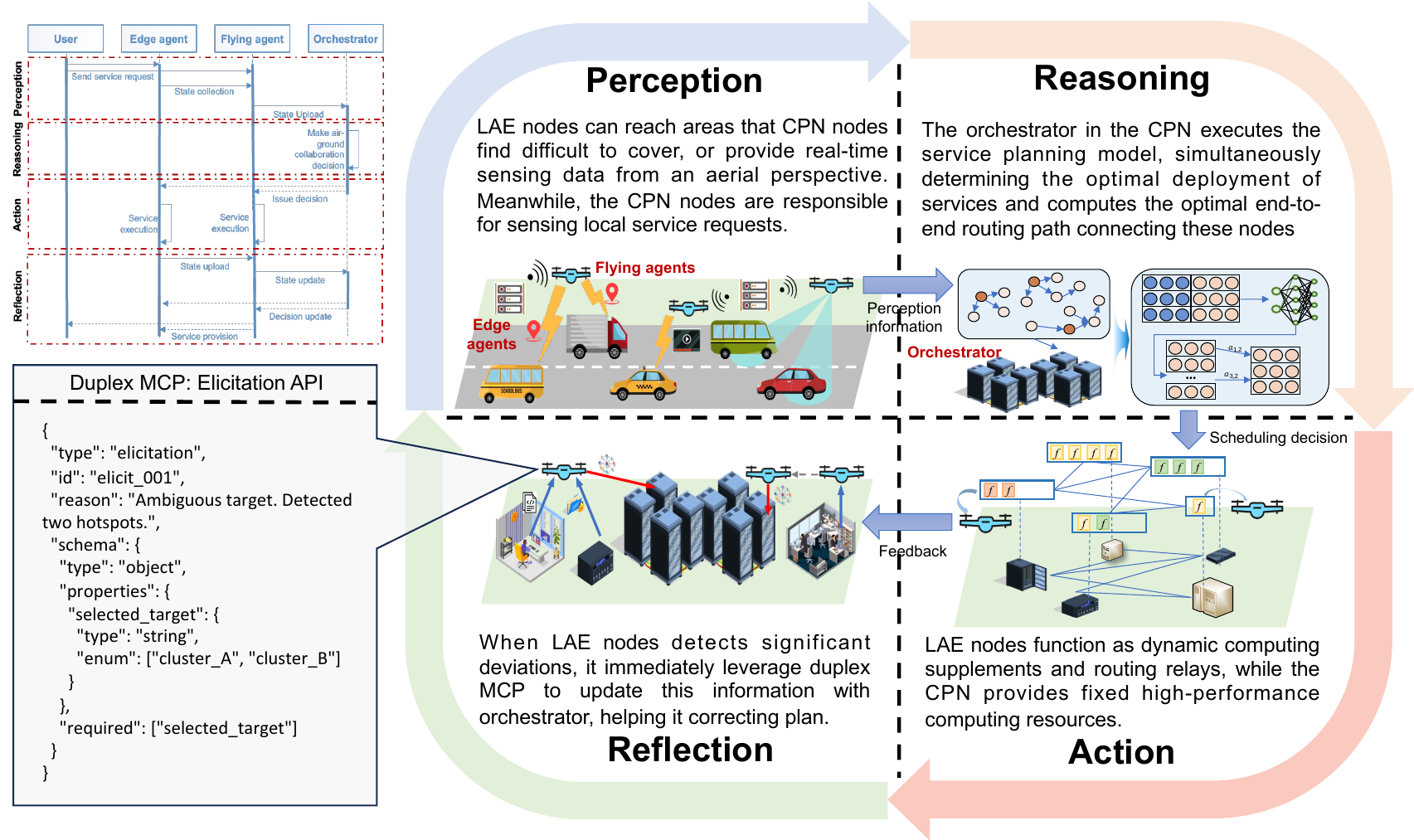}
\caption{The agentification approach constructs an architecture that integrates perception, planning, action, and reflection capabilities. By treating both LAE and CPN nodes as autonomous, collaborative intelligent agents, this paradigm breaks down the operational silos and inherent communication friction between these two heterogeneous systems.}
\vspace{-4mm}
\end{figure*}

To address this challenge, we introduce the agentification paradigm\cite{ref17}, transforming the rigid service orchestration model into a flexible and intelligent collaborative ecosystem. As shown in Fig. 2, the agentification approach constructs an architecture that integrates perception, planning, action, and reflection capabilities. This paradigm enables LAE and CPN nodes to operate as a unified entity.
Rather than functioning in isolation, they act as intelligent agents capable of communicating with each other, sharing real-time state information, and collaboratively executing tasks. Specifically, it transforms the traditional one-way rigid control into a flexible bidirectional collaboration. This agent-initiated real-time state aggregation enables the CPN-based service orchestrator to efficiently update its global situational awareness and trigger on-demand re-planning. Based on the updated global view, the orchestrator can dynamically schedule mobile LAE nodes to move toward hotspot regions, thereby achieving real-time air–ground resource coordination and effectively addressing dynamic service hotspot problems.
\subsection{System Overview}
As shown in Fig. 2, our proposed agentification architecture is composed of three primary elements: 
\begin{itemize}
    \item \textbf{CPN Centralized Orchestrator}: It is deployed at the core of the CPN and is primarily responsible for the Reasoning phase illustrated in Fig. 2. The orchestrator manages the connections of all execution agents, receives perception information, such as environmental information and network information, from them, executes the service planning model, and computes optimal end-to-end service routing paths that interconnect these nodes.
    \item \textbf{Edge Agent}: The edge agents in CPN provide stable computing resources and handle local service requests. In the Perception phase, these nodes sense and report local service requests. In the Action phase, the edge agents execute the orchestrator’s scheduling decisions by providing fixed, high-performance computing resources.
    \item \textbf{Flying Agent}: Flying agents are jointly formed by the LAE node clusters. As mobile and reconfigurable network entities, LAE nodes significantly enhance the CPN's flexibility and coverage. During the Perception phase, they leverage their mobility to provide real-time aerial sensing data from areas inaccessible by CPN nodes. In the Action phase, they serve as dynamic computational supplements and mobile routing relays. During the Reflection phase, they are responsible for detecting significant environmental deviations—such as the two service hotspots identified in the API example on the left side of Fig. 2—and immediately reporting them back to the orchestrator.
\end{itemize}

Meanwhile, we employ the Model Context Protocol (MCP) into the proposed agentification framework. MCP is a protocol specifically designed for agentic systems, serving as an interface that connects AI models with the external world. In this architecture, the orchestrator located at the core of the CPN acts as the MCP Host. It treats all heterogeneous execution agents as MCP Servers.

\subsection{Duplex MCP-enabled Agentification Architecture for Air-Ground Collaboration}
To efficiently facilitate the bidirectional communication between CPN and LAE networks, we design a duplex MCP-assisted agentification framework to enable two-way invocation between agents and the orchestrator. In this framework, we integrate an elicitation mechanism into the MCP layer of execution nodes, allowing LAE or CPN nodes to pause ongoing operations and proactively query the orchestrator for additional information or confirmation during task execution. Due to dynamic network conditions, the orchestrator’s initial plan can become obsolete during execution. Duplex MCP converts this rigid one-way paradigm into a closed-loop interaction, enabling LAE agents to actively invoke the orchestrator's MCP interface and provide real-time updates, achieving better hotspot response.

Our Duplex MCP-enabled agentification architecture extends the standard MCP protocol stack. In traditional frameworks, the interaction model is primarily strictly top-down. Although agents can transmit data back to the orchestrator through reverse RPC or similar techniques, such interactions are typically passive state reports. These frameworks lack a mechanism that allows an agent to proactively interrupt the orchestrator’s execution flow and request re-planning, which is a capability that becomes critical when the agent detects significant local environmental changes, such as shifts in service hotspot regions. The core of our Duplex MCP design specifically addresses this missing coordination-and-correction mechanism. In our design, the core of Duplex MCP lies in the integration between the agent’s internal reflection module and its state machine. During task execution, the reflection module continuously monitors both the prerequisites of the assigned plan and the real-time environmental data, such as the link quality of an LAE agent or the computational load of a CPN node.
When the reflection module detects significant deviations, such as an unexpected surge in load or a disrupted critical link that renders the original plan infeasible, it immediately triggers the state machine to transition into an Elicitation state.

In this state, the agent pauses its current operation and, via its MCP client module, actively initiates an active call to the orchestrator. This call encapsulates critical contextual information, including agent metadata, associated service information, and abnormal data observed. The orchestrator, continuously listens for such agent-initiated requests. Upon receiving one, it parses the contextual data and updates its global situational awareness model accordingly. This update triggers an immediate re-planning process, wherein the orchestrator recalculates the service plan and issues an updated command through the standard MCP interface. The agent then receives the new plan, transitions back from the Elicitation state to the Action state, and resumes execution, thereby completing a closed-loop, adaptive interaction cycle.

\subsection{Numerical Results}
We implemented the above agentification architecture on a server equipped with an AMD EPYC 7763 CPU and 4 GeForce RTX 4090 (24GB) GPUs. We uniformly deployed 20 edge computing nodes and 20 LAE nodes within a square area to simulate three environments: a pure CPN environment, a pure LAE network environment, and a CPN–LAE integrated environment\footnote{Our codes are available at https://github.com/Hollowday/lae-cpn}.

\begin{figure}[t]
\centering
\includegraphics[width=3in]{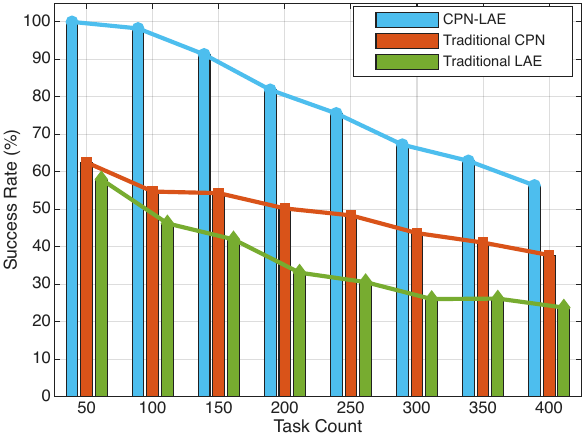}
\caption{Task success rate versus task count}
\vspace{-4mm}
\end{figure}
In our experiment, we first ignore service hotspot effects and simulate random task requests sent to random nodes. A task was considered failed if its waiting time exceeded a certain threshold. As shown in Fig. 3, the task success rate decreases as the number of concurrent tasks increases in both the LAE and CPN networks. The decline is particularly pronounced in the LAE network due to its insufficient computational capacity to handle large-scale concurrent tasks. Although the CPN provides ample computational resources, it lacks flexibility. When integrated with the LAE network, LAE nodes can assist computing nodes by performing partial data preprocessing and extending the CPN’s coverage, thereby achieving a higher overall task completion rate.

\begin{figure}[t]
\centering
\includegraphics[width=3in]{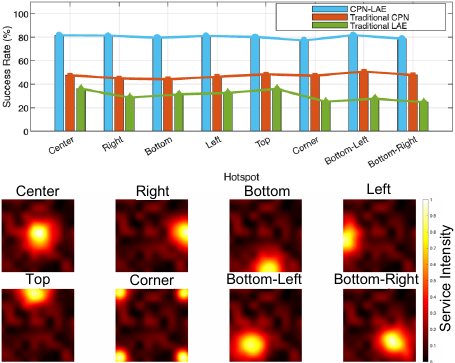}
\vspace{-0.3cm}
\caption{Task success rate versus movement of hotspot area}
\end{figure}

Next, we demonstrate the advantages of the proposed CPN–LAE agentification framework in handling dynamic service hotspot variations. In this experiment, we fix the total number of concurrent tasks while gradually increasing the proportion of service requests originating from a specific hotspot region in each round, and observe the resulting changes in service success rate. As shown in Fig. 4, when service requests become concentrated in one region, both the CPN and LAE networks exhibit a significant drop in success rate. In contrast, under our framework, the orchestrator can leverage the situational awareness provided by LAE nodes to dynamically dispatch them toward hotspot regions, where they act as relay nodes and offload part of the computation to less-loaded CPN nodes in nearby areas, thus achieving a higher overall service success rate. Moreover, the Duplex MCP mechanism enables LAE nodes to immediately report hotspot changes to the orchestrator, allowing it to perform real-time scheduling adjustments and maintain system stability under highly dynamic load conditions.

\section{Future Directions}

\subsubsection{Digital Twin-enhanced LAE-CPN Integrated System} Digital twins can create virtual models of the LAE-CPN infrastructure for pre-deployment simulation, validation, and optimization. This enables proactive resource management and predictive maintenance to enhance network performance.

\subsubsection{Security and Privacy in LAE-CPN Integrated System} LAE-CPN integration introduces vulnerabilities such as exposed LAE nodes and an expanded CPN attack surface\cite{ref1,ref3}. 
Future research requires robust, multilayered security, including lightweight encryption for UAVs, privacy-preserving techniques for distributed data, and real-time intrusion detection.

\subsubsection{Energy-aware LAE-CPN Integrated System} Energy efficiency is critical due to LAE's battery limits and CPN's high power use\cite{ref1}. Future work should design energy-efficient mechanisms, like intelligent task offloading that balances latency and energy, and dynamic routing protocols to optimize UAV trajectories and communication.

\section{Conclusion}
In this paper, we have explored the intrinsic synergy between LAE and CPN, and have proposed an integrated air–ground computing and communication architecture. We have analyzed the complementary characteristics of these two paradigms: LAE can provide mobility, environmental awareness, and flexible network coverage, while CPN can offer powerful global computing orchestration, reliability, and intelligence. Through a case study, we have shown that our integrated approach achieves a significant improvement in task success rate, particularly when handling dynamic service hotspot scenarios. Finally, we have outlined key future research directions.

\bibliography{reference}
\bibliographystyle{IEEEtran}

\end{document}